\documentclass{svjour2}

\newcommand{\ket}[1]{\ensuremath{|#1 \rangle}}
\newcommand{\bra}[1]{\ensuremath{\langle #1|}}
\newcommand{\braket}[2]{\ensuremath{\langle #1|#2 \rangle}}

\newcommand{\iden}{\ensuremath{{\sf 1\hspace*{-1.0ex}\rule{0.15ex}
{1.2ex}\hspace*{1.0ex}}}}

\begin{document}
\title{An introduction to many worlds in quantum computation}
\author{Clare Hewitt-Horsman}
\date{}
\maketitle



\begin{abstract} The interpretation of quantum mechanics is an area of increasing interest to many working physicists. In particular, interest has come from those involved in quantum computing and information theory, as there has always been a strong foundational element in this field. This paper introduces one interpretation of quantum mechanics, a modern `many-worlds' theory, from the perspective of quantum computation. Reasons for seeking to interpret quantum mechanics are discussed, then the specific `neo-Everettian' theory is introduced and its claim as the best available interpretation defended. The main objections to the interpretation, including the so-called ``problem of probability'' are shown to fail. The local nature of the interpretation is demonstrated, and the implications of this both for the interpretation and for quantum mechanics more generally are discussed. Finally, the consequences of the theory for quantum computation are investigated, and common objections to using many worlds to describe quantum computing are answered. We find that using this particular many-worlds theory as a physical foundation for quantum computation gives several distinct advantages over other interpretations, and over not interpreting quantum theory at all.\end{abstract}

\section*{Introduction}

Recent years have seen a small, but steady, increase in interest surrounding the interpretation of quantum mechanics. There is a growing awareness that previously the boundaries of what can be investigated in physics have been drawn too tightly, and that questions that have formerly been rejected as meaningless may in fact correctly be asked. Perhaps the most important of these questions concern what exists. What, for instance, are the physical processes of quantum mechanics? What is the physical structure of the universe? The different ways of answering these questions gives rise to the different `interpretations' of quantum mechanics. The aim of this paper is to give an introduction to one such interpretation, a `many worlds' theory. This will be done from the specific point of view of quantum computation; anecdotally, this is a field in which curiosity about these questions has been relatively strong. This presentation will, however, be accessible (and, hopefully, interesting) to those working within any area of quantum mechanics. I will not assume any previous acquaintance with issues involved in interpreting quantum mechanics, and will keep the exposition as free from technical terminology as possible. Readers with a background in this area may therefore find some of the discussion to be a little circuitous but not, I hope, confusing. A `suggestions for further reading' section is included at the end for any readers wishing to follow up on the material presented here.

\section*{Interpreting quantum mechanics}

Why do we need to interpret quantum mechanics? This is probably the most widespread immediate response to the enterprise. After all, we have had nearly a century of dazzling discoveries, both theoretical and technological, courtesy of the formalism of quantum mechanics, none of which seem to require any commitment to a given interpretation. The `shut up and calculate' method has become so standard that very often any deviation from it is viewed with suspicion. Where, then, is the need for an interpretation?

To begin to answer this, let us ask ourselves a couple of questions about quantum mechanics. The first question to consider is: why is it so successful? Why does the formalism (plus the Born rule) work so well at predicting the results of experiments? This is a fairly basic question, and the fairly obvious answer is that it is a true theory. Quantum mechanics works because it is correct: within its limitations (a necessary caveat as we do not have a quantum gravity) it is right.

So what do we mean when we say that quantum mechanics is true? What is it that makes it true? The temptation at this point is to say that what we mean by true is correct: it correctly predicts future behaviour. Unfortunately this is not a very good answer to our original question, as we have put ourselves in the position of saying that quantum mechanics works because it works! We know it \emph{does} work; what makes this the case?

This is actually another basic question to which a basic answer can be given: quantum mechanics is true because of the way things behave in the world. Quantum mechanics accurately represents the way the world works: atoms etc (or some other kind of `stuff') move around and interact in such a way that the quantum formalism can be used to predict what will happen to them. In short, quantum mechanics is true because that's the way that physical reality is set up.

Let us look at a simple example of what we mean here. Consider a black box containing some electronics (configuration unknown) with a switch and a light. Certain ways of toggling the switch will make the light blink in various ways. After experimenting with this box we come up with a method of predicting what patterns of blinks follow which inputs. This is a true theory of how the box works, and is true because of how the things in the box work. If we were to open it up we would see a circuit that controlled how the output responded to the input. Because of the way this circuit in the box behaves, our theory of input and flashes is correct; put another way, what determines whether our theory is true or not is the configuration of electronics in the box.

These two fairly trivial questions, why quantum mechanics works and why it is true, lead us, rather surprisingly, to a non-trivial conclusion. If the world is set up such that quantum mechanics is true, then that quantum mechanics \emph{is} true can in itself tell us something about the set-up of the world. If quantum mechanics were not true, then physical reality would have to be different. So the fact that we find that it \emph{is} true gives us information about the world. In our box example, if the box responded to inputs in a different way then the electronics inside would have to be configured differently. So how the inputs and outputs relate tells us things about how the electronics are put together. Indeed, were we knowledgeable enough about electronics then we might even be able to deduce the circuit structure from those alone, without having to open the box. Moving back to quantum mechanics, our trivial questions, therefore, lead us to ask very significant ones: what must physical reality look like in order for quantum mechanics to work? What is the physical structure of the world? What sort of physical things exist, and how do they behave?

It is these questions that are addressed by an interpretation. An `interpretation' of quantum mechanics is a physical theory of how the world works such that quantum mechanics gives us correct predictions. All of these theories give the same results as the standard formalism for current observations and experiments, but some of them give different predictions in certain (usually extreme) situations. For example, some hidden-variables theories modify the Schr\"{o}dinger equation slightly. Some interpretations also make predictions that are simply outside the scope of the formalism. For example, in some dynamical collapse theories it takes a certain amount of time for the wavefunction to collapse at measurement, time which in theory is measurable. All of this is very much `in theory': at the moment we do not have the experimental capability to distinguish between interpretations.

This does not, however, mean that we have no way of choosing between interpretations in order to find the best one, that is the one that most accurately reflects reality. We always have more criteria than experimental results with which to chose between our theories. To take an extreme example, let us return to our black box. Suppose I am very good at electronics and can work out the circuit without opening the box. As an alternative `interpretation' of the box I can postulate the existence of a pink candy-floss daemon sitting in the box flashing the light on and off whenever the switch is pressed. If we cannot open the box then we cannot tell the difference between these two theories experimentally. We would not, however, send an article into Physical Review citing the pink candy-floss daemon as the explanation for the box's behaviour. For fairly obvious reasons, demonic candy-floss (of any colour) is a worse explanation of what happens inside the box than the circuit explanation. It is much less simple than the alternative: we would have to then explain the existence of the creature, how it was made, when and how candy-floss became sentient, how it got into the box, why it is giving that output for that input, etc. It also requires us to add to our collection of things that exist: the circuit theory needs only the existence of electronics (to which we already -- presumably -- subscribe), whereas the alternative requires pink candy-floss daemons to exist as well. So we go with the best explanation for the behaviour of the box: the circuit.

Although no-one has yet advanced a pink candy-floss interpretation of quantum mechanics, we can use the same criterion to select between the ones that we do have: which is the best explanation of the observed phenomena (that the quantum formalism works)? We will be looking in detail here at one particular interpretation, a `many worlds' or `Everett' style theory. We will concentrate on this one as it is the best of the available interpretations, for reasons that will be given. Another, better, interpretation may of course come along later and supercede it. That, however, is a possible fate for any physical theory: we can only ever choose between the theories that we actually have. At any given time, it is our best available theory that we want to look at. If we want to know how the world is, it is that theory that we ask. That theory might turn out to be wrong in the future, but at a given time it is the best guide that we have to the way the world is set up, and we are entitled to follow it.

\section*{Introducing many worlds}

A good way to introduce the main ideas of the many worlds interpretation is to look at what is called the measurement problem. As we all know, quantum mechanics predicts undetermined states for microscopic objects most of the time: for example, in an interferometer the photon path is indeterminate between the two arms of the apparatus. We deal with such states all the time, and are seemingly happy with them for the unobservable realm.

Such happiness is destroyed when we consider an experiment (such as the infamous Schr\"{o}dinger's Cat set-up \cite{silly}) where macroscopic outcomes are made dependent on microscopic states. We are then faced with an `amplification of indeterminism' up to the macro-realm: the state of the system+cat is something like
\begin{equation} \ket{0} \otimes \ket{\mathrm{cat \ dead}}+ \ket{1} \otimes \ket{\mathrm{cat \ alive}}
\label{cat} \end{equation}

This is the measurement problem: how do we reconcile the fact that quantum mechanics predicts macroscopic indeterminism with the fact that we observe a definite macro-realm in everyday life?

Almost all proposed solutions to the measurement problem start from this assumption: that a superposition of states such as (\ref{cat}) cannot describe macroscopic physical reality. In one way or another all terms bar one are made to vanish, leaving the `actual' physical world. The exception to this way of solving the problem was first proposed by Everett \cite{everett}. His interpretation has since been modified and improved, but the central posit remains the same: that physical reality at all levels is described by states such as (\ref{cat}), and each term in such a superposition is equally actualized. 

Dispute over what this actually \emph{means} gives rise to the myriad of different Everettian interpretations that we have (Many Worlds, Many Minds, etc. etc. etc.). One thing we can say about all Everett interpretations is that they include at some level and to some degree a multiplicity of what we would commonly refer to as the `macroscopic world': each term in the superposition is actual, the cat is \emph{both} dead \emph{and} alive.

Even before we go further than this, there are two pressing problems here for the Everettian. Firstly, there is the logical problem: how can anything be in two states at once? Secondly, we have the measurement problem itself: if all terms are real, why do we only see one? Looking at the first problem, we note that we do not get a logical contradiction when something is in two different states with respect to an external parameter. For example, I am both sitting at my desk typing and standing by the window drinking tea, with respect to the parameter time: \emph{now} I am at my desk, but (hopefully) \emph{soon} I will be drinking tea. The parameter in Everett theories with respect to which cats, etc., have their properties is variously called the world, branch, universe, or macrorealm. The idea (at a very basic level) is that in one world (branch, etc.) the cat is dead, and in another it is alive. Extending this to include experimenters we get the answer to our second question: in one world the experimenter sees the cat dead, in another she sees it alive.

We now have the problem of making these rather vague ideas concrete. As noted above, the differing ways of doing this give rise to different Everett-style interpretations. We shall now turn to a specific theory (chosen as the best of the Everett-style theories on offer), an amalgam of the ideas of Everett \cite{everett}, Saunders \cite{saunders,saundersrel}, Vaidman \cite{vaidman} and Zurek \cite{zurek,zurekdande}, and the expansion of these by Wallace \cite{david,davidstr} and Butterfield \cite{butterfield}, which we will call the neo-Everettian interpretation\footnote{I am indebted to Harvey Brown for this moniker.}.

\section*{The neo-Everettian interpretation}

The main ideas of the neo-Everettian interpretation are the following. The totality of physical reality is represented by the state $\ket{\Psi}$: the `universal state'. There is no larger system than this. Within this main structure we can identify substructures that behave like what we
would intuitively call a single universe (the sort of world we see around us). There are many such substructures, which we identify as single universes with different histories. The identification of these substructures is not fundamental or precise - rather, they are defined for all practical purposes (FAPP), using criteria such as distinguishableness and stability over time, with decoherence playing an important role such an identification. 

The main structure is often termed the `multiverse' to distinguish it from the many `universe' substructures. In each of these universes in turn we can find smaller substructures which are in general more localized than an entire universe. These are known as \emph{worlds}. For example, we would describe (\ref{cat}) as referring to worlds of the Schr\"{o}dinger cat apparatus, without reference to the state of the rest of the universe/multiverse. An important aspect of worlds is that they are, in general, stable over only certain time-scales. It is not the case that if we can identify certain worlds at certain times then we will necessarily be able to identify them at all subsequent times. We will see that the important point is how long we want to be able to identify the worlds for: if they are stable over those time-scales then we can use them.

The main mechanism from which we gain this stability of worlds is decoherence. It is the linchpin of the neo-Everettian response to the measurement problem,  allowing the stable evolution of definite substructures of universes within the universal state. We will not here go through all the mechanics of the decoherence process (this may be found in many places, for example \cite{Zeh} and \cite{omnes}), but merely state the relevant points. One interesting result of this use of decoherence that we will see is the explanation of why measurement has been so important in quantum theory. Measurements generally decohere the system being measured, by coupling them to a large environment. Measurement is therefore important as one way in which decoherence happens; it is also important to note, however, that this removes the idea of measurement as a \emph{fundamental} concept in quantum mechanics. 

Decoherence occurs when a system becomes entangled with a larger environment. If we then look at the behaviour of the system alone then we have to trace out the environment, which leads to the loss of the interference terms between states of the decoherence basis\footnote{The decoherence basis for large objects is one in which position and momentum can be (approximately) well defined, and which is stable over long time-scales -- a `classical-like' basis.}. Thus, at any given instant, we can identify a multiplicity of non-interfering substructures, which are worlds. Furthermore this lack of interference persists over time if decoherence is ongoing: that is, individual substructures (elements of the decoherence basis) evolve virtually independently, with minimal interference with other such substructures.

We now find ourselves with the problem of precisely how we are to define these worlds. This is perhaps the part of a many-worlds interpretation that is the hardest. For example, in a na\"{i}ve many worlds interpretations (eg \cite{deutsch1985}) one cannot find the preferred basis to specify the worlds using decoherence as their structure is absolute and decoherence, as is well known, is only an approximate process: we get a specification of branches for all practical purposes (FAPP), but not in any absolute sense. It is a common assertion in the literature on the preferred basis problem that the preferred basis must be defined absolutely, explicitly and precisely, and that therefore the failure to give such a definition (and indeed the impossibility of doing so) is a terminal problem for an Everett interpretation.

By contrast, in the neo-Everettian interpretation the worlds structure is \emph{not} absolute, and so no such explicit or precise rule is required\footnote{This vital understanding is found in \cite{davidstr}, from which the material for this section is taken.}. The key to understanding how this works is to move away from the idea, implicit in much of the literature, that the measurement problem must be solved at the level of the basic theory: that is, that we must recover a single macrorealm (or the appearance of one) in the fundamental posits of our theory. The neo-Everettian theory does something different, by defining the worlds, in Wallace's phrase, as `higher order ontology'. The structures of the worlds and branches are not precisely defined, but FAPP they are a useful way to look at the overall state. Furthermore, we as observers are some such structures, and so we must look at the evolution of these structures and the rest of the state relative to them in order to recover predictions about the world we live in from quantum mechanics -- which also gives us the answer to the measurement problem.

Physics (and indeed science in general) is no stranger to the idea of using approximately-defined structures. In everyday life we deal with them all the time. For example, we can go to the beach and (if we are in a suitably meditative mood) count the waves as they come in. If we are feeling more energetic then we can paddle out and use one particular wave to surf on. The waves exist as real entities (I can count them and surf on them), they persist over time as distinct structures (we can follow them as they come into shore and break), and if I surfed on one then I would talk about \emph{the} wave I caught.

Waves are, however, not precisely defined: where does this wave end and that one begin? Where does this one end and the sea begin? Different water molecules comprise it at different points in its history -- given this, how is the wave defined? We cannot find any method that will tell us absolutely when a given molecule is part of the wave or not, and this is not merely a \emph{technical} impossibility: there is simply no fact of the matter about when a wave ends and the sea begins. We can use rough rules of thumb, but at the fine-grained level there is no precise fact to find.

We thus see that there are many objects that we would unhesitatingly call real that we nevertheless cannot define absolutely and objectively. Such entities are part of our higher order ontology, not `written in' directly in the fundamental laws of the theory, but nevertheless present and real and explanatorily useful.

It is at such a level that the neo-Everett concept of a world operates. It is not an entity written into the fundamental laws of the interpretation: in fact, what neo-Everett does is (merely?) explain how, from quantum mechanics alone (decoherence is a straightforward consequence of the basic laws), the structures can emerge that describe the world that we see around us everyday\footnote{This is in fact one of the great strengths of neo-Everett as an interpretation: there are no mathematics added to standard quantum mechanics (a strength particularly for those physicists who do not wish a theory to be changed for conceptual or philosophical reasons); it is truly an \emph{interpretation}.}. These structures are not (and cannot be) absolutely defined, but this is no argument against their reality.

The standard neo-Everettian approach, which stops here, leaves us with something of a problem as regards quantum computing. If worlds are these entities defined by decoherence, which therefore do not interfere to any great extent, then a many-worlds analysis of a quantum computation simply says that there is one world, in which the computation is happening, and that is it. Not unreasonably, we might wonder why we should bother with the interpretation at all. What we can do, however, is to extend the standard approach into the domain of quantum computation, using the principles that we already have to give us a many-worlds view in a setting of coherent states. In this we are definitely departing from the standard neo-Everettian approach but not, I would argue, by very much.

The fundamental principle of the neo-Everettian approach is that all parts of the state are real. Most of the time we prefer to talk about the decomposition of the state into worlds because that is what we are familiar with: one particle has one spin, one computation step computes one value, etc. How we perform this decomposition is entirely up to us. Usually we prefer worlds that do not interfere very much with each other, and which preserve this independence and are stable over quite long time-scales. However, the notion of recognizing familiar patterns within a state can be extended into the situation where that state is coherent. The time-scale over which these patterns will persist will be much shorter than that of worlds given by decoherence -- they may, indeed, be \emph{de facto} instantaneous. However, if they are useful then we are entitled to use them. 

Defining worlds within a coherent state in this way is a simple extension of the FAPP principle that has been described above. If our practical purposes allow us to deal with rapidly changing worlds-structures then we may. As we are dealing with coherent states, the worlds-structures will in general be subject to interference over the time-scale of an operation, and the `relevant time-scales' over which worlds are defined will be smaller than that of the single operation. This is not, however, a real difference from situations in which decoherence defines the worlds, as even then we have to deal with the (albeit generally theoretical rather than practical) possibility of decoherent systems re-cohering.

So in order to use the neo-Everettian approach for quantum computation we are extending the set of circumstances in which a `world' is defined. This is in line with the underlying motivation of the neo-Everett approach, in which we identify familiar patterns within a state that are stable and independent over relevant time-scales. The relevant time-scales which we will use will be defined entirely FAPP -- and can include instantaneous time-scales. Such objects, though, remain `worlds': they are the familiar objects of a decoherent system \emph{over the relevant time-scales}.

This fits in well with intuitions that are often expressed about the nature of quantum computations, especially those based on the quantum Fourier transform \cite{qft} and quantum walks methods of computation \cite{qw}. There are frequently statements to the effect that it \emph{looks like} there are multiple copies of classical computations happening within the quantum state. If one classical state from a decomposition of the (quantum) input state is chosen as an input, then the computation runs in a certain way. If the quantum input state is used then it looks as if all the classical computations are somehow present in the quantum one. We will go into greater detail later on about the nature of computation under a many-worlds picture, but for now we will simply say that the recognition of multiple worlds in a coherent states seems both to be a natural notion for a quantum information theorist, and also a reasonable notion in any situation where `relevant' time-scales are short.

\section*{Challenges to neo-Everett}

Now that we have seen in detail what the basics of the neo-Everettian approach are, it is time to deal with common objections, both to a many-worlds view of quantum mechanics in general, and to the neo-Everettian version in particular. These are what is known as the ``incredulous stare" argument, and the problem of probability, which is arguably the biggest problem faced by the Everettian of any stripe.

\subsection*{The incredulous stare}

\begin{quotation}\emph{Entities should not be multiplied beyond necessity}.\end{quotation}

\noindent This is the oft-cited piece of advice that, traditionally, William of Ockham gives us on constructing our physical theories. One of the most common objections to any many-worlds theory is that it violates Ockham's Razor by massive multiplication of entities (\emph{ie} worlds). This objection can take the form of simply saying that one cannot \emph{really} be serious in thinking there is such a mind-bogglingly huge number of worlds. At a rather more sophisticated level, the objection is that such a huge increase in what we are committing ourselves to believe exists cannot but tell against the theory.

Things are not, however, this simple. We do not think that one theory is better than another simply because one commits us to less `stuff' than the other. Modern cosmology tells us that the universe is so big that ``you just won't believe how vastly, hugely, mind-bogglingly big it is,"\footnote{As an aside, this must surely be an objection to the simple `incredulous stare': given that the enormity of a single universe creates boggling, in what sense is the boggle produced by many universes that much worse? And why is one boggle acceptable and the other not?} \cite{HHGG}, and yet we much prefer it as a cosmological theory to the Aristotelian universe that ended just beyond Saturn. The same may be said for atomic theory: rather than accepting that, say, there is one table here, we must accept that there are vast numbers of atoms and sub-atomic particles stuck together -- not only that there is more `stuff' than an alternative table-only theory, but that there are more \emph{kinds} of stuff than the alternative. Another example is dark matter: this is the postulation again of large quantities of a completely different type of `stuff', and yet this is not generally considered to be a fatal flaw in the theory.

The clue to all this lies in Ockham's Razor itself -- entities are not be multiplied \emph{beyond necessity}. We do not choose between theories simply by looking at which theory postulates the fewest entities or types of entities. If the entities are \emph{necessary} then we are entitled to have them in our theory. So what makes an entity `necessary' in this context? An entity becomes necessary if it is given by the best explanation of the observed phenomena that the theory is trying to account for. Our best theories of cosmology include a huge universe containing vast quantities of matter -- and so we accept its existence, multiplying our entities enormously, but not beyond necessity. The same is true for atomic theory and dark matter, although the latter gives us some additional insights into the process. Dark matter is postulated because we don't want to give up on such a large set of successful theories. In order to keep those theories intact, we need to add not only to the amount of `stuff' in the universe, but also to the number of types of objects there can be. This is considered by many to be a price well worth paying.

So we see that the simple objection to an increase in entities only works if they are not necessary. If they do enough work for us, then we can keep them -- there is a trade-off between introducing more `stuff' and the use that it can be put to in the theory. It is only when all other things are equal between theories that the one with the fewest entities is to be preferred. And all other things are \emph{not} equal when we look at the interpretations of quantum mechanics. By introducing many worlds into our set of things that exist, we can give the best current explanation for the phenomenon that quantum mechanics works. This claim will now be justified. This is not the place to go into the details of alternative interpretations of quantum mechanics -- the interested reader is directed towards the `suggestions for further reading' at the end of the paper. Rather, we will concentrate on the differences between the alternatives and the neo-Everettian interpretation, and why it is a better theory.

There are two main contenders with the neo-Everett theory for best explanation of the way the world is such that quantum mechanics works. These are the various types of hidden variable theories such as \cite{bohm1952}, and the dynamical collapse models \cite{grw}. In hidden variables theories, particles etc really do have single, definite, values for quantities such as momentum and position (the eponymous `hidden variables'), it's just that we can never find out what they are. All we have access to is the probability distribution, from which quantum mechanics is constructed. The hidden variables can either be fixed, or there can be a probability distribution over them; in either case, the Schr\"{o}dinger equation is modified. Dynamical collapse models take `the collapse of the wavefunction' to be a real, dynamical, process and adjust the mathematics of quantum theory accordingly. 

The first thing to note is that, unlike the neo-Everettian theory, both hidden variables and dynamical collapse models require a change to the formalism of quantum theory. This is not something to be accepted lightly. The formalism of quantum mechanics is extremely successful, and we would need very good reasons to change this. Moreover, what is being done is not truly \emph{interpretation}: the question with which we started this paper was: given that quantum mechanics is correct, what is the world like? A hidden variable or dynamical collapse theory is positing a different version of quantum mechanics, rather than using the one that we have -- and that has stood up to nearly a century of testing -- to find out what the world is like. So the question that must be asked is: are many worlds so unthinkable that we are willing to change the quantum formalism in order get rid of them? If we stick with quantum mechanics as we have it, then the only option is many worlds.

Suppose we were to reject standard quantum mechanics. What would we get in return? The answer, in both cases, is a much more complicated theory. In both hidden variables and dynamical collapse theories there is the addition of an extra layer of complex formalism on top of the standard quantum theory. Because this is being suggested as fundamental, there is no hope that the basic theory may become simplified. Not only do we have less simplicity in our formalism (which is bad enough), but both these alternative theories also bring in their own commitments to additional entities. Various different hidden variables theories add different things, from `corpuscles' to carry the hidden variables, to quantum potentials to push them around. Dynamical collapse theories bring in a `thing' that is the wavefunction that can collapse, and is interacted with. None of these entities behave in a way that we are entirely familiar with -- so we are committed to a significant increase in the number of types of thing that exist, and hence to a more complex system of interactions between them. An additional level of complexity comes in when we consider the hidden variables theories -- not only do we have new entities behaving in new ways under new laws, but there is an absolute bar on us ever being able to see them. There seems to be no better explanation of this given than that it must be there or else quantum mechanics would not work.

So if we want a single-world quantum theory then we are going to pay dearly for it. Our formalism will become much more complicated, without the expectation that a more fundamental, and simple, form may be found. We have to introduce new kinds of entities, entirely different from anything else that we think exists. Moreover, these entities act and interact in a completely alien fashion; and, unlike the case of dark matter, this doesn't even enable us to keep our old theories. The alternative is significantly simpler. In return for the acceptance of an increased number of entities in the world, we can have a simple fundamental formalism, a simple fundamental dynamics and no increase in the different types of things or interactions that we think exist. One can hardly conclude from this that entities are being multiplied beyond necessity.

It is worth re-iterating at this point that we cannot simply say that we will reject \emph{all} current interpretations of quantum theory and hold out for one that is single-world but does not have all the drawbacks of current single-world theories. We can only choose from theories that we have, not between current theories and something that does not exist. We should also bear in mind that, despite decades of trying, no-one has managed to come up with a single-world interpretation without severe costs. An argument can be made that, to a certain extent, it is not even possible that such an interpretation could exist: \emph{somehow} the multiple outcomes of the standard formalism must be made reduce to one, which must involve additional formalism or entities, or both.

Even from this initial discussion we can see that the pay-off from accepting an increase in number of worlds is large, and the argument could even be made at this stage that a many-worlds theory is evidently the best theory for quantum mechanics. We will find as we proceed that there are even more arguments in favour of the neo-Everettian interpretation, and we will discuss them in detail as we reach them. Before turning to these further advantages, however, we will discuss what is widely perceived to be the greatest disadvantage of any many-worlds theory: the infamous problem of probability.

\subsection*{Probability}

Let us turn again to the Schr\"{o}dinger's cat experiment. Suppose we leave the diabolical device for the half-life of the radioactive atom, giving a probability of $\frac{1}{2}$ that the cat is dead. In a many-worlds picture, the cat is both dead and alive, one state in each world. Now, what happens if we leave the device for longer, say for tens of half-lives? We still have two identifiable patterns in the state, one with a dead cat and one with an alive cat. What has changed within the neo-Everettian world-view? Why is it that in one scenario it is much more probable that the cat is alive than in the other, when it seems that our physical theory says that these are identical scenes?

This appears at first sight to be a fatal problem for the neo-Everettian approach. Unlike other many-worlds theories, we cannot even resort to the dubious expedient of counting worlds to tell us which outcome is more likely. The theory simply puts a measure over the worlds, given by the modulus-squared of the wavefunction, and leaves us with no idea of how that measure translates into probabilities. This is the problem of probability.

In recent years, work on this problem has largely centred around Deutsch's 1999 proposal \cite{deutschprob} to use decision theory to try to show that a rational agent in an Everettian universe would lay bets on outcomes according to a measure that is identical with the modulus-squared of the wavefunction. The argument sometimes appears rather as a conjouring trick: an apparently solid result is produced from seemingly innocent assumptions as if from thin air. As such it is often, and I think rightly, viewed with suspicion. 

In its recent, improved, form \cite{dwdt} the argument begins by considering how a rational agent would act if they knew the universe to be Everettian. Standard assumptions from decision theory are combined with the centrally-important postulate of \emph{equivalence}: if two outcomes have the same quantum-mechanical amplitude, then they should be given the same likelihood by the rational agent. With these assumptions and postulates in place, the distribution of likelihoods by the agent is easily (and fairly uncontentiously) shown to follow the Born rule for the modulus-squared of the wavefunction.

Not unreasonably, objections (and arguments) have centred on the postulate of equivalence. The simplest objection is that the postulate begs the question, smuggling in the very connection between amplitudes and rational likelihood (and hence probability) that we are attempting to ground \cite{barnum}. Such an objection, however, ignores the large amount of work that goes in to arguing for the postulate: none of the proponents of the decision-theoretic programme claim that equivalence should be self-evident.

The arguments in favour of equivalence take the form of a demonstration that no other way of giving likelihoods to outcomes with equal amplitudes is rational. Various quantum ``games'' are analysed to show that the rational agent bets on equal-amplitude outcomes as if they were were equally likely. Moreover, any other bets would be in fact \emph{irrational}: there is only one way to define the relative likelihoods of two outcomes with identical quantum amplitudes.

These arguments have attracted a great deal of counter-argument, and remain deeply contentious. Objections have been raised to the completeness of the discussion into alternative likelihood distributions based on similar-amplitude outcomes \cite{plewis}. Other objections centre on the description of physically distinct games by the same amplitude for the outcome, arguing that concentrating on the part of the wavefunction that we already think of as ``the probability of the game outcome'' begs the question of what a rational agent should take as a likelihood \cite{hprice}. Further objections can be made to the programme as a whole: even if quantum amplitudes and rational likelihoods can be shown numerically identical, what does that tell us about probability? The fact that two quantities are the same does not necessarily tell us what the physical and conceptual connections between them are\footnote{We should be particularly wary of this in quantum mechanics, where we have the cautionary tale of the original measurement problem. This may be described as the difficulty accounting for the numerical identity of a decohered pure state and a mixed density matrix (so-called proper and improper mixtures).}. An even wider objection could be made to the use of decision theory itself in a multiple-outcome quantum theory, that in order to define the decision-theoretic concepts in such a universe, the same conceptual work would need to be done as to define probability itself. The ``derivation'' of probability from decision theory would therefore be a curiosity without much real content.

It is certainly true to say that opinion remains deeply divided as to the success of the decision-theoretic programme. It is, however, important to point out that this way of solving the problem of probability is not an integral part of the neo-Everettian theory; the acceptability or otherwise of the theory does not turn on the adoption or rejection of the decision-theoretic argument.

The reason why we can take this view is that, in fact, probability in neo-Everett is not nearly as problematic as it seems at first glance. Or rather, we should say that it \emph{is} problematic -- but only as much as it already is, both for any theory of quantum mechanics and also for everyday life. The problems with probability that face the neo-Everettian interpretation are no worse than those facing any theory that attempts to deal with probabilities other than 1 or 0.

To see what we mean by this, let us return yet again to our unfortunate feline. Let us abandon the idea of multiple outcomes, and say that Schr\"{o}dinger's cat will be either dead or alive at the end, and that's it. Again let us run two experiments, one with an end state given by $ \frac{1}{\sqrt{2}} ( \ket{dead} + \ket{alive} ) $, and the other by $ \frac{1}{\sqrt{3}} ( \sqrt{2}\ket{dead} + \ket{alive}) $. We discover, happily, that at the end of both experiments the cat is alive. Now how are we to understand this? In both cases we have the same physical outcome: the cat is alive. In one scenario we want to say that this is the more \emph{likely} outcome, but what is the physical difference? It seems that we simply have a measure over somehow `potential' outcomes, with no idea how that measure translates into probabilities. And what could a probability possibly mean anyway? Either the cat turns out dead or alive, so how can we tell if one outcome was more likely than the other? What does it mean for us to say, before the box is open, that a tragic outcome is more or less likely than a happy one? And if we were gambling on the outcome, why would it not be entirely rational to place equal bets in the two cases?

Of course we all know how to use probabilities in everyday life, including everyday quantum mechanics, and I am not for a moment advocating abandoning the whole notion of probability because we don't know how to ground it, any more that I would argue for us to give up arithmetic because we do not, fundamentally, know what makes ``2+2=4'' correct. The important point at issue, though, is whether the adoption of a neo-Everettian theory makes our understanding of probability even worse than it already is. I argue that all it in fact does is show up the existing problems of probability by demonstrating them in a different situation. 

So far, it seems we are equally in the dark about probabilities whether we have a single- or multiple-outcome theory. There is, however, a standard way in which physics often tries to make sense of probability. This is what we are all taught as statistical mechanics: probabilities refer to properties of ensembles that are large enough to be considered, for our purposes, infinite. So, for example, $ \frac{1}{\sqrt{2}} ( \ket{dead} + \ket{alive} ) $ tells us that, when a large number of cats are prepared in this state, roughly half will be found upon examination to be dead, and half alive. This is, of course, very useful for statistical mechanics where we have large numbers of particles that can act as our ensemble (although even there we encounter problems as it isn't \emph{actually} infinite). However, it does not seem to get us very far with our two experiment runs. It's not particularly enlightening in either case to be told what the result of a large number of identical runs would be, as we are only running each experiment once. Do we then have to say that for a single run (or a small number of runs) there is no such thing as probability? So are our two experiments in fact identical? Alternatively, given that actually we would quite like to say that there is such a thing as ``single-shot'' probability, we might say that the probabilities are really given with reference to what \emph{would} happen \emph{if} a large number of identical experiments were run. However, now we are in the uncomfortable position of wishing to call probability a \emph{physical} property of a \emph{hypothetical} ensemble. It is difficult to see how such a definition could be meaningful, let alone such a thing possible.

A further problem for those seeking to criticize neo-Everett on probability grounds is that we do, of course, have ensembles and frequencies of outcome in neo-Everettian worlds. Within each world there will be access to exactly the same notions as a single world in a single-outcome theory. So even if an understanding of probability in terms of ensembles could be given, that understanding would translate straight over into the neo-Everett interpretation\footnote{It is of course true that in some neo-Everettian worlds that if, say, a fair coin were tossed a million times, then the frequencies of heads and tails would not match the limiting distribution. However, this is something that we have no problem thinking is possible(!) in a single world, so the mere existence of worlds such as this is not an argument against neo-Everett.}.

It is, then, perfectly allowable simply to say that we don't know what probabilities mean in the neo-Everett theory, as we do not know what they mean in any other theory. If we are willing to use probabilities at all, then we should be willing to use them in neo-Everett, as it does not bring in any different problems -- it simply presents the same ones in a different light. Probability problem are not unique to multiple-outcome interpretations of quantum mechanics. If the arguments of Deutsch et. al. from decision theory to probability are accepted, this does not simply bring the neo-Everettian interpretation up to the standard of everything else; rather, it would be a tremendous breakthrough in the understanding of probability, and would be a very powerful argument for the truth of the theory. Contrary to popular perception, then, probability is no problem for neo-Everett. At worst it is neutral, being neither better nor worse than other theories, but at best it is an immense bonus of the theory, and provides a compelling reason to think it true. It remains, however, an area of controversy. We will turn now to what is, uncontroversially, a major advantage of the neo-Everettian interpretation, and one that is an important part of the argument that this interpretation is the best available for quantum mechanics: locality.






\section*{Locality}

In the experience of the author, a common reaction to the disclosure that the neo-Everettian interpretation is local is bemusement, and then an increasing suspicion that one is perpetrating a long and involuted joke. A slightly less common reaction is simply to assume that one of the periodic arguments for a loophole in the Bell inequality experiments is being made. In the light of this, we will start with the important facts about the neo-Everettian approach and locality and then go on to show how they come about. Firstly, it is local, and local in the way that we normally mean when we talk about locality -- we are not getting our locality by the back door by changing the meanings of the term. Secondly, neither the Bell inequalities nor the corresponding experiments from Aspect onwards are considered to be incorrect or to have loopholes\footnote{I am not claiming that these cannot exist, nor am I taking up a position on, for example, recent work on the meaning of the Bell inequalities \cite{joychristian} -- I am simply saying that these are not necessary to the locality of neo-Everett.}. These claims will, of course, strike many as impossible, so we will now investigate how they can possibly be true. It will be useful at this point to give a sort of sneak-preview of the result: essentially, the neo-Everettian approach can be local because separated systems don't need to signal in order to demonstrate that they are entangled, as all possible outcomes of a measurement are realized.

It is an incorrect, but unfortunately widespread, belief that the physical world has been shown experimentally to be nonlocal in character. The chain of this argument starts with the original paper by Einstein, Podolsky and Rosen (EPR) \cite{EPR}, carries on through Bell \cite{bell1}, and finishes off with the Aspect experiments \cite{aspect}. We will now consider this argument from the point of view of the neo-Everettian interpretation. Because issues of nonlocality are notoriously complex, we will go into these arguments in some detail. 

\subsection*{The EPR paper}\label{epr}

The form of locality put forward in the EPR paper is known now as `Einstein locality': no causal influences of any sort are allowed between spacelike separated objects. This is sometimes contrasted with `Bell locality', where only currently known causal influences are considered. Einstein locality requires that at all levels of a physical theory there can be no causation happening outside the lightcone, that there is an absolute meshing with the spacetime structure of relativity theory, not just with its phenomenology. This is usually argued for by appeal to paradox: were we able to signal outside our lightcones then it would be possible to send signals back in time and, for example, prevent those same signals being sent in the first place. However, Einstein locality goes beyond the requirement that such paradoxes cannot be constructed, and states that even causal influences that cannot be used to signal in this way cannot propagate between spacelike separated objects\footnote{These constructions of Einstein locality come later, in explications of the EPR paper: in the paper itself this form of locality is assumed.}. 

The argument of the EPR paper concerns the \emph{completeness} of quantum theory (that is, whether ``every element of the physical reality... [has] a counterpart in the physical theory"), and is given in two parts. Apart from the criterion of reality, the first part is not actually necessary for their arguments. This criterion says that 
\begin{quote} If, without in any way disturbing a system, we can predict with certainty (ie with probability equal to unity) the value of a physical quantity, then there exists an element of physical reality corresponding to this physical quantity. \end{quote}

In the second part of the paper, EPR assume that quantum mechanics is a complete physical theory and then show that this entails the contradiction that it is \emph{not} complete. They do this by looking at the reality of values of non-commuting operators.

Two particles are entangled and then sent to different locations. If an observable $ \hat{X} $ is measured on the first system a then the state of the particle b is given by $ \psi_x^{(b)}$. Similarly, if an observable $ \hat{Y} $ is measured on a then the state of b is given by $ \psi_y^{(b)} $. Now if there is no signalling at any level between a and b (this is the place where the locality criterion enters the argument), and if QM is a complete theory, then $ \psi_x^{(b)} $ and $ \psi_y^{(b)} $ describe the \emph{same} reality, that of the actual physical state of b.

EPR now consider the case where $ [X,Y] \neq 0 $. In this case, the two wavefunctions belonging to the same reality are eigenfunctions of incommeasurable operators, and from this EPR conclude that the values of these operators are real simultaneously. A way of seeing how this works is the following. In the real world we measure $ \hat{X} $ on a. There is however a \emph{possible} world in which we measure $ \hat{Y} $ on a. In the actual world, b then has a value of $ \hat{X} $ predictable with certainty. In the possible world, b has a value of $ \hat{Y} $ predictable with certainty. However, any values that b has must be the same in both this actual and possible world because there is no causal link between a and b, so which one of $ \hat{X} $ and $ \hat{Y} $ is actually measured cannot make a difference to the values possessed by the particle b. Therefore, b has actual values of $ \hat{X} $ and $ \hat{Y} $. The strength of this argument lies in the nature of the reality criterion: it depends on whether something can be \emph{predicted} with probability 1, rather than depending on any actualization.

The idea that these operator values have simultaneous reality is, naturally, in defiance of the initial assumption that QM is complete. They therefore conclude that QM is not complete\footnote{What we have used here is not the logic of the paper as it is presented. EPR use the first part of the paper to show, formally, $ [p \vee q] $ where\\
p: quantum mechanics is not complete\\
q: values of noncommuting operators do not have simultaneous reality \\
They then use the second part to show $ [\neg p \rightarrow \neg q ] $. Put together, they conclude $[p \vee q ] \wedge [\neg p \rightarrow \neg q ] \vdash p.$ However, in the course of the second part they prove, as we have shown, the much simpler proposition $ \neg p \rightarrow p \vdash  p $, again giving them the desired conclusion that quantum mechanics is not complete.}. 

\subsection*{Bell}\label{bineq}

We can, of course, negate the EPR conclusion if we drop the locality principle. In this case, then, the measurement performed on system a affects the real state of affairs at b and the EPR argument fails. The EPR paper therefore leaves us with a choice: either quantum mechanics must be non-local in a sense that causes the EPR argument to fail, or else it is incomplete. If it is incomplete then it must be completed -- which would lead to a hidden-variables type theory.

The most fully worked-out hidden variables theory is the de Broglie-Bohm theory (see for example \cite{bohm}), and it was this theory that Bell was interested in when he came up with his famous inequalities \cite{bell1,bell2}. In the de Broglie-Bohm theory the wavefunction is defined on the configuration space of particles, and changes to one particle have nonlocal effects on other particles via the quantum potential. Bell wondered if this was a feature that \emph{any} hidden-variables theory must have, and came up with his inequalities as a way of answering this question. If the inequalities were violated by quantum mechanics, then no local hidden variables theory could reproduce its results.

We will start by considering the particular type of hidden variables theories to which the de Broglie-Bohm theory belongs: that of deterministic hidden variables theories. The values of the hidden variables exactly determine the outcome of measurements with no residual probabilistic behaviour. Consider a very simple Bell-like set-up. The spin of an electron, $ e_1 $, is measured along the directions $ \mathbf{a} $ or $ \mathbf{a}^\prime $, and that of a second electron $ e_2$
is measured along $ \mathbf{b} $ or $ \mathbf{b}^\prime $, resulting in measurements $ a_n, a^\prime_n, b_n, b^\prime_n $ (each of which is $ \pm 1 $). The Bell function is then
\begin{equation} \gamma_n = a_nb_n + a_nb^\prime_n + a^\prime_nb_n - a^\prime_n b^\prime_n \label{bfunc} \end{equation}
\noindent A Bell inequality can only be derived if the RHS can be factored; that is, if
$$ \gamma_n = a_n (b_n + b^\prime_n) + a^\prime_n (b_n - b^\prime_n)$$

This is the locality principle of a deterministic hidden variables theory. The conditions under which this is possible are similar to EPR's locality principle: that the results that occur when $ e_1 $ is measured are not contingent on what is being measured on $ e_2 $. That is, that $ a_n $ and $ a_n^\prime $ are the same regardless of whether $ b_n $ or $ b_n^\prime $ was being measured on $ e_2 $. 

Since Bell, work has also been done on a more general class of hidden variable theories, that of stochastic hidden variables theories. In these cases, the hidden variables do not fully determine the results of experiments, but rather determine a probability distribution for the outcomes. This changes the conditions under which a Bell inequality can be derived. The locality principle here cannot deal with individual outcomes, as in the deterministic case, but rather must concern the probability distributions that the hidden variables set up. The condition for a Bell inequality to be derived in this case is known as \emph{factorizability}. If we have two sets of possible results at $\mathbf{a}$ and $\mathbf{b}$, $\{a_i\}$ and $\{b_i\}$, and a distribution of hidden variables $\lambda$, then factorizability means that
\begin{equation} P(a_i, b_i, \lambda) = P(a_i | \lambda). \ P(b_i | \lambda) . \ P(\lambda) \label{fac}\end{equation}

Following \cite{shimony2}, this condition is usually decomposed further into two separate conditions, both of which must hold for factorablity to be possible. They are \emph{parameter independence} and \emph{outcome independence}\footnote{In \cite{jarrett} these are termed ``locality" and ``completeness".}. Parameter independence states that a probability distribution cannot be affected by manipulating causally unconnected entities (such as the settings of a remote apparatus). That is, if the $\{\lambda\}$ are fixed, then changing the remote apparatus setting does not affect the probability distribution. Outcome independence states that, for a fixed set of the hidden variables, any pair of outcomes of measurements on unconnected systems must have independent probabilities.

The combination of outcome independence and parameter independence forms what is known as `Bell locality' (see for example \cite[p75]{redhead}).

\subsection*{Statistical locality}

As well as the EPR and Bell results, there is a third important constraint on the nature of quantum non-locality. This is that, if any such non-locality exits, then it cannot be used to communicate non-locally. This is owing to the well-known no-signalling theorem (see for example \cite{zandb}), which states that information cannot be transmitted by manipulating quantum correlations. A user-friendly proof of this can be found in \cite[p116]{redhead}, which we will outline here.

There are two systems, A and B. System B can be subject to a perturbation acting only on itself, $ \iden \otimes U_B $. $ a\otimes \iden $ is an operator acting on system A only. Working in the interaction picture, $ a $ evolves under the unperturbed Hamiltonian and the overall wavefunction changes only when a perturbation is present. We model an attempt to change the statistical properties at A by a perturbation on B.

\noindent The wavefunction after such a perturbation (at time $ t^\prime $) is given by
$$ \ket{\psi(t^\prime)} = (\iden \otimes U_B) \ket{\psi(t)}$$

\noindent The time-evolved operator $ a $ at $ t^\prime $ is $ a(t^\prime) $. We now consider the statistical distribution of $ a $ after the perturbation has been applied:
\begin{eqnarray*} \bra{\psi(t^\prime)} (a(t^\prime) \otimes \iden) \ket{\psi(t^\prime)} & = & \bra{\psi(t)} (\iden \otimes U_B^{-1})  (a(t^\prime) \otimes \iden )(\iden \otimes U_B) \ket{\psi(t)}\\
{} & = & \bra{\psi(t)} (a(t^\prime) \otimes \iden) \ket{\psi(t)} \end{eqnarray*}

\noindent That is, the probability distribution at A is the same regardless of any perturbation on B. This is known as statistical locality, because it shows that for two systems which are only connected by quantum correlations (rather than causal influences) the statistics at one cannot be affected by operations on the other. It is important to note that, unlike the Bell and EPR arguments, this is a straightforward mathematical theorem with only one prior assumption: that the probability distribution is given by $ |\braket{\psi}{\psi}|^2 $. This being such a foundational part of quantum mechanics, it is safe to say that the conclusion that statistical locality must be obeyed is an extremely strong one.

One immediate consequence of the no-signalling theorem is that Einstein locality can never be violated at a phenomenological level\footnote{See \cite{chrisharvey} for a full discussion of how QM is phenomenologically compatible with relativity.}. Different theories about how quantum statistics come about may have non-local elements in them (as for example hidden variables theories do) or be structurally incompatible with relativity, but the statistics themselves can never show a violation of Einstein locality. Whatever non-locality there is, it must be hidden to this extent.

\subsection*{Local realism}\label{locrela}

We now have what looks like a complete argument for non-locality in QM. After deriving his inequalities, Bell showed that in certain situation QM \emph{does} violate the inequalities, so any hidden variables theory must have non-local elements. Together with the EPR argument this seems clear-cut. Either QM is non-local or it is incomplete, but even if it is incomplete then it must still be non-local (non-local here in the Bell sense of violating parameter or outcome independence (or both), rather than the stronger Einstein sense). However, this non-locality cannot be used to create anything like causal paradox because it cannot be used to send information. Such a view of locality in QM is correct for almost all realist interpretations. For each interpretation the definition of locality is slightly different (see \cite{redhead}), but fundamentally the notion is the same, as is the fact that there is non-locality (albeit hidden). 

Very often the language of ``local realism" is used to describe the state of affairs that is to be ruled out by this chain of reasoning\footnote{See for example the original Aspect paper \cite{aspect}, which was entitled ``Experimental Test of Realistic Local Theories via Bell's Theorem".}. The EPR argument for the incompleteness of QM (and hence the necessity for a hidden variables theory) can be circumvented both by denying the (Einstein) locality principle (as is the case with, for example, dynamical collapse theories) or the realism principle (the Copenhagen interpretation, as local and non-realist, is a good example \cite{bohr}). As a consequence, it is often incorrectly concluded that the only possible interpretation of quantum mechanics which could incorporate both realism and locality is one in which quantum mechanics as we have it is incomplete - that is, a hidden variables theory. The Bell inequalities, as conditions that possible local hidden variables theories must satisfy, are then seen as the conditions on a local realist interpretation. As they find that such an theory cannot reproduce the results of quantum mechanics, it is common to find statements such as \cite{santos} that describe tests of the Bell inequalities as ``quantum mechanics vs local realism".

Such statements are incorrect. They do not follow from the arguments given above because there is a third way in which the EPR argument can fail, which does not necessitate the denial of the assumptions of either locality or realism. The third assumption that can be questioned is that which makes the contradiction right at the end of the EPR argument: that if the values of $\hat{X}$ and $\hat{Y}$ have simultaneous reality but not simultaneous predictability then quantum mechanics is incomplete. This is only a contradiction if there is only a single world in the theory. In the case of a many-worlds theory, the values of $\hat{X}$ and $\hat{Y}$ have simultaneous reality in different worlds, but the non-predictability of values of non-commuting operators pertains to within a \emph{single} world. In such an interpretation, the real and possible worlds of the exposition above are both actual worlds. 

It is interesting to note that as we have broken the argument at the EPR stage, as far as Everett theories are concerned the Bell inequalities are not in the first instance connected with questions of locality. `Bell locality', in the forms of parameter and outcome independence, does not necessarily mean any such thing -- they are conditions that must be met for factorability to occur, and this is a locality condition dealing specifically with hidden variables theories. It does not necessarily mean anything for Everett theories -- and, indeed, outcome independence is \emph{prima face} violated whenever there is entanglement.

It is therefore not necessary that a local realist theory is a hidden variables theory -- the EPR argument does not rule out the possibility that a many worlds theory could be local and realist. Hence, without an argument that this is not possible, we are not entitled to describe the Bell inequalities as debating `local realism'. They deal with the conditions on one type of local realist theory, but not all types. If a similar phrase is wanted, `single-outcome local realism' describes what is being discussed -- for every interaction there is only a single outcome and hence a single world. Only within that restriction does the EPR argument show that a local realist theory must be a hidden variables one. It is perhaps the best-kept secret in quantum mechanics that the Bell inequalities do not track locality when there are multiple outcomes of experiments, even though the literature on this goes back more than two decades (\cite{page,stapp,rubin1,rubin2,guido,chrisharvey}, amongst others). Indeed, it is difficult to understand in this situation what the elements of a Bell function such as (\ref{bfunc}) could refer to: $a_n$ etc. are defined as single outcomes to single experiments, in a way that we do not have in the neo-Everettian interpretation. If we look further at the factorizability criterion (\ref{fac}), we can see even more clearly that this refers to entities that simply are not present in our theory: we have neither hidden variables $\lambda$, nor classical probability distributions over single outcomes. These are all elements of the hidden-variables theories to which Bell inequalities refer, so it is not entirely surprising that in their absence, in an Everett-type theory, the inequalities tell us nothing about the locality of the theory.

So we are left with the possibility that `multiple-outcome' local realism is possible. This is of course not to say that any many-worlds theory is necessarily local. Most, in fact, are not -- for example, Deutsch's original version of his many-worlds approach had instantaneous non-local splittings of worlds. However, the neo-Everett interpretation with which we have been dealing is a local realist theory that reproduces exactly the results of orthodox quantum theory.

The realist credentials of neo-Everett are obvious, dealing firmly as it does with what the theory says exists `out there', but its locality may be slightly less evident. Moreover, this is strong, Einsteinian, locality with no causation outside the lightcone. We know from the EPR argument that if  an Everett theory is local and realist, the fundamental reason for this must be because it is multiple-outcome, as this is the only other way to evade the argument. That is, when a measurement (or indeed any other interaction -- measurement is not special in neo-Everett) occurs, it is not required that one outcome out of many is chosen to be special. For example, in an EPR-type experiment most interpretations of QM need a mechanism whereby the second particle knows what was measured on the first, in order for the corresponding state to be actualized. Because \emph{all} possible states of the second particle are actualized anyway in neo-Everett, this sort of a `signal' is not needed.

Of course, the second particle is still correlated with the first in neo-Everett, and at first glance it would seem that some sort of non-local signal \emph{is} needed, to tell the worlds of the first particle how they join up with the worlds of the second particle. This is indeed necessary in those many worlds theories that have spatially extended worlds that split instantaneously. It is not, however, necessary for a faithful many worlds interpretation to have such a worlds structure, as there is also the fact of statistical locality. Correlations between distant systems cannot be seen while the systems are still separated -- it is only when information about the statistics from both systems is brought together in the same place that the correlations can be calculated. What happens in neo-Everett is that, instead of spatially extended worlds from the outset, when the systems are separated they have worlds local to those systems, with no connection to the separate system. When the information on the systems is brought together, however, the two separate sets of worlds `join up' to make one set. If there is any correlation between the two systems then the worlds join up in a specific way; if not, the joining is completely random. Therefore spatially extended worlds only occur when the relevant systems have come into causal contact, such causation being entirely physical and within the light-cone. No non-local effects of any kind occur either to create the worlds or within the worlds. 

To see how this works in detail, consider the simple Bell experiment above. We will look only at the worlds that are relevant to the experiment -- as noted before, there are many other possible ways of decomposing the state. Particle $ e_1 $ is measured along $ \mathbf{b} $ and $ \mathbf{b}^\prime $ $ n $ times, and $ e_2 $ is measured along $ \mathbf{a} $ and $ \mathbf{a}^\prime $ also $ n $ times. Spatiotemporally local to $ e_1 $ there are many worlds, corresponding to all the possible outcomes of $ 2n $ measurements, and similarly for $ e_2 $. So, for example, if $n=1$ then at $e_1$ we have four worlds. In one there is an electron that has been measured along $\mathbf{a}$, a detector registering ``$a_1 = +1$", and (if we like) an experimenter who is reading this detector, and who sees this single outcome to a measurement along $\mathbf{a}$. In the other worlds we have the similar situations for the other three outcomes. However, at this point there is no way for an experimenter at either site (and in any world) to construct a Bell inequality -- they have access only to their own statistics which, because of the no-signalling theorem, are not dependent on the experiments happening in the other place. It is only when, for example, the experimenter on $ e_2 $ sends her results to $ e_1 $ that he can then construct an inequality and see the correlations. It is at this point that the worlds corresponding to measurement result on $ e_2 $ come into contact with those of $ e_1 $ and, in a completely local operation, join up as dictated by the entanglement between the two particles. So in the end the experimenter who sees a certain set of statistics at one site joins up with the experimenter at the other site who sees statistics which are correlated due to entanglement with the first set.

It is the same for all quantum phenomena that exhibit what are normally called `non-local' effects. It is only when information has been communicated from one to another \emph{via} a classical channel that the `non-locality' becomes manifest. Another good example is superdense coding\footnote{See \cite{chrisharvey} for a full discussion.} \cite{nandc}: although at first sight it looks as if Alice has changed the state of Bob's qubit when she performs operations on her own, it is only when she then sends Bob her qubit that he can extract the information from the correlations between their qubits. A further example can be found in \cite{mevdh}, where a fully formalized entanglement swapping protocol is worked through locally.

It is difficult to overstate the importance of this result. To begin with, there is the significance for quantum mechanics itself -- that a faithful, local and realist physical implementation of the formalism is possible. This has major implications for all aspects of quantum theory, and for quantum computation in particular. It is not uncommon to find nonlocality referred to as a computational resource, closely connected with entanglement, which is seen as its generator. Entanglement is widely regarded as the particularly \emph{quantum} resource available for computing, and is indispensable in an analysis of quantum computing. If it is no longer to be viewed as fundamentally non-local, that opens up the possibility of being able to track a `flow' of information during computation, showing how information is transmitted and stored. This also leads to the possibility of a local analysis of quantum computation, and indicates that we can locate the information processing at every point in particular qubits or qubit groups\footnote{This was the motivation behind the development of the Deutsch-Hayden formalism \cite{dh,mevdh}. More broadly, this `logical Heisenberg picture' \cite{dan}, while not an integral part of the neo-Everettian interpretation, nevertheless does good work as a native formalism for the theory. Locality is manifested by the assignment of a `descriptor' to each system, the properties of which change only under local operations. This formalism also demontrates the formal as well as dynamical separability of the interpretation: the descriptors for subsystems completely determine the full system descriptors. If we are happy with calling the descriptors \emph{properties} of the subsystems (physically meaningful even though they are not locally fully determinable), then local properties fully determine global ones. }. This is particularly interesting in the case of distributed quantum computing \cite{distqc}, where teleportation is used to link quantum processors. Not only will an analysis of the processors themselves be local, but we will also be able to track the information flow between processors as it is carried by the bits in the `classical channel' of teleportation. 

For quantum theory more generally, the most important advantage of locality is that it removes one of the major barriers to forming a theory of quantum gravity. There is no longer the necessity of reconciling a structurally nonlocal quantum mechanics with a fundamentally local spacetime -- quantum theory can now be interpreted as itself fundamentally local . On a more theoretical level, within quantum theory itself we also have the removal of the somewhat unsatisfactory situation that there is some manner of `conspiracy' that produces a non-locality that is nevertheless even in principle uncontrollable (the no-signalling theorem). With no non-locality, statistical locality does not appear as if by magic, and the mathematical agreement with relativity emerges from a structural similarity.

This, then, is one of the most important arguments for the neo-Everettian interpretation. If we accept many worlds, in the neo-Everettian form, then we can have a local quantum mechanics. No other realist quantum theory gives us this, as any single-outcome realist interpretation will be subject to the EPR+Bell argument. Locality is so important, both for information processing and for the creation of a quantum theory of gravity, that there must be significant drawbacks with any theory that offers it, if we are not to accept that theory. I would argue that the acceptance of multiple outcomes is a small price to pay for a local quantum mechanics.



\section*{Computation in many worlds}

The use of the neo-Everettian interpretation has many implications that are specific to quantum computing, and in this section we will discuss some of the main areas. We will then show how the neo-Everettian theory rebuts objections that have previously been raised to the whole notion of describing a computation in many-worlds terms. At the same time these objections, combined with some interesting new developments in the field, will help us see the most important consequenses of a neo-Everettian view in computation.

Along with locality, probably the most important consequence of a neo-Everettian view of computation is the relation between quantum and classical processors. Put simply, in neo-Everett there is no fundamental quantum/classical divide. Bits are not, in fact, a different sort of thing altogether from qubits. Classical processes are a subset of quantum processes, in specific circumstances. This is an advantage on two fronts. Firstly, the fact that we do not have a distinct idea of where the classical/quantum divide is anyway is explained by there not \emph{being} a distinct divide. Secondly, we now have fundamental connections between the different resources of computing. Rather than two separate sets of resources that nevertheless may be interchanged in certain situations\footnote{For example, with shared entanglement, two bits may transmit one qubit of information (teleportation) or \emph{vice versa} (superdense coding).}, we now have physically the same resource being put to different uses -- a bit is a qubit with the coherence data neglected. One direct advantage of this is that we can formalize an entire protocol using the same method, rather than requiring a formal shift between bits and qubits at various points.

This picture also changes our ideas of how information is processed. Again, it is not that there is a fundamental difference of kind between quantum and classical computation; rather, classical computation is a subset of quantum. The difference becomes one solely of the conditions under which the computation happen, such as decoherence. One particularly fascinating consequence of this is that in both cases, of quantum and of classical computations, there are multiple computational worlds. This is one of the biggest changes in how we have to view classical computing, as decoherent quantum computing; the difference is in the presence or absence of the coherence data.

As a consequence, questions that are often asked about the relationship between quantum and classical computing need to be turned on their head. Instead of starting from the view that we know what classical computing is, that it is all sorted out and understood, we begin from \emph{quantum} computation as basic. It is not quantum computing that is `strange' and in need on explanation, but rather classical -- rather than ask why quantum computers can do more than classical, we ask why classical computers are restricted from carrying out all the tasks of a quantum computer. 

In a similar way, our understanding of bits and qubits needs to be reversed. Rather than attempting, as has historically been the case, to define qubits in terms of bits, in our neo-Everettian picture it is the bits that stand in need of description. Rather than being defined in terms of classical messages, a bit becomes the information transmitted from a decoherent qubit. 

This, then, is the picture of computation that neo-Everett gives us. Quantum information, processing and communication is fundamental, and the classical counterparts must be constructed by giving a set of restrictions to the quantum situation. It is those restrictions that are responsible for the difference in computational ability of quantum and classical systems. All processing and communication is local and continuous; information does not `jump' between coders or processors, it must all be transmitted by a physical system moving between points in accordance with relativity. Fundamentally, all parts of the state of a computation exist, both in the classical and quantum cases.



\subsection*{``A Quantum Computer Only Needs One Universe''}

We can see already that there are significant advantages to considering quantum computation from the point of view of neo-Everett. The view does, however, have its detractors. We will consider here in detail the paper \cite{steane}, which presents several arguments against the use of Everett-style theories to understand quantum computing. Dealing with the points raised will not only show how the arguments are not valid for the neo-Everettian theory given here, but will help fill out the conception of quantum information theory and computation that we have. 


The main argument of \cite{steane} is that ``[q]uantum computation is... not well described by interpretations of quantum mechanics which invoke the concept of vast numbers of parallel universes. ", and that statements along the lines of ``a quantum computer can perform vast numbers of computations simultaneous" are ``sufficiently misleading that [they] should have a `health warning label', ". I will address Steane's seven `remarks' and show why I think that, contrary to a `warning label', a neo-Everettian description of quantum computation should come with a `glowing recommendation' label! It is important to note here that I am not making the claim that Deutsch gives \cite[p217ff]{deutschfr}, that quantum computing is \emph{proof} of a many-world theory. We don't need to make such a claim that quantum computers are, in Steane's words, ``wedded to `many worlds' interpretations" in order to see that neo-Everett has distinct advantages as a physical picture for computing and, I would argue, is the best physical explanation for the phenomena. Furthermore, we will see that some of the elements of the discussion that have been taken to be indicative of a single-world viewpoint do actually emerge naturally from our neo-Everettian theory.

\subsubsection*{Objections from Information Theory}

Steane's three remarks, 1, 2 and 6, object on the basis of information theory. In the first remark he notes that the information content of the output of a quantum computation is the same as for a classical one of the same length. Steane thus uses this to say that it is ``not self-evident that a quantum computer does exponentially more computations than a classical computer". 

We first note that we do not, typically, define the complexity of a calculation in terms of the information content of the output. To take an extreme example, the output of both of the two questions ``does 1+1=2?" and ``can every even number greater than 2 can be written as the sum of two primes?" is one bit (yes/no), yet we would not wish to say that the resources needed to calculate them are the same! Secondly, it is important to realize that this way of framing the question takes classical computation as basic, in the exact way that was warned against above. A quantum computer is not constructed by `gluing together' many classical computers -- we say instead that within a quantum computation we may identify many computational worlds. If we are to model quantum computers in classical terms then of course we will need exponentially many more computations happening in a given time-step; however, in the neo-Everettian picture this is not a modelling in fundamental terms. We only need to explain how a quantum computer can perform exponentially more calculations if we take classical computing as the basic point of reference. The neo-Everettian theory does not, so is not touched by this argument.

This is, in fact, precisely what Steane goes on to warn us about in his second remark: that we should not base our definitions of computational complexity on what is possible classically. We are therefore in complete accord on this point.

Remark 6 is an interesting argument from the point of view of neo-Everett. Steane argues that more efficient quantum versions of classical algorithms do not generate as many intermediate (classical) evaluation results -- as he puts it, ``extraneous classical information". Putting his argument in neo-Everettian terms, this can then be used to argue against being able to identify multiple classical-style computations within the main state. This in fact shows up the problems that can arise when a classical understanding is taken as primary, because it is not true that such ``extraneous classical information" is never generated, but rather that we do not have the technological ability to extract it. Consider a quantum algorithm based on the quantum Fourier transform, consisting of a global Hadamard gate\footnote{A Hadamard gate transforms a basis into the orthogonal basis. Under a Hadamard gate, the computational basis transforms as $\ket{0} \rightarrow \frac{1}{\sqrt{2}} ( \ket{0} + \ket{1} ) \ ; \ \ \ket{1} \rightarrow \frac{1}{\sqrt{2}} ( \ket{0} - \ket{1} )$. The importance of this gate in an algorithm is that the initial $\ket{\mathbf{0}}$ state is thereby transformed into an equal sum of all possible computational states.}, a global function gate\footnote{This acts as $\ket{\alpha}\ket{0} \rightarrow \ket{\alpha} \ket{f(\alpha)}$ where $f(x)$ is the function that the gate evaluates.}, and then another global Hadamard. During the middle,  manipulation, stage, suppose the state of the register is measured to give the value of $f(\alpha)$. In the neo-Everettian picture, we are left then with multiple measuring devices, each registering a particular value, and multiple experimenters looking each at this piece of `classical' information. Usually, by now decoherence would have removed the phase information from the vicinity of the experimenter, and each measuring device and person would be quantum mechanically separate. However, suppose we had the technology to manipulate the measuring devices and experimenters\footnote{If it is considered that a person brings in unnecessary complications then she may be replaced by a computer.} quantum mechanically, and to shield them on long time-scales from decoherence with the outside world. In this case, all the separate worlds can be re-interfered, and the algorithm continued, incorporating the measuring devices etc.. In each world there would have been a measurement of the classical information content, and so a generation of the classical intermediate results affecting the wider world, but because of the shielding from decoherence this did not prevent the algorithm being completed. We can see in this example the lack of a fundamental quantum/classical divide in neo-Everett: even classical information may still be used in a quantum situation.



\subsubsection*{Mathematical Notation}

In his third remark, Steane makes the point that often mathematical notation can be
misleading, and that there are some cases where one can interpret (na\"{i}vely) what has happened as many more
processes than have actually occurred. Thus there is no straight argument from the existence of decompositions of
the state such as (\ref{cat}) to the existence of many worlds.

Steane is correct that notation can be misleading, and also that we can sometimes incorrectly say that many
processes have happened when only one has. However, one of the ideas that has hopefully been demonstrated in this paper is that there is more going on in the neo-Everett picture than simply
extrapolating from the existence of the decomposition (\ref{cat}) to the existence of many worlds. As we have
said, what defines the worlds is their explanatory usefulness and their stability and independence. Were these
criteria not fulfilled for the states in the decomposition, then we would have no right to call them
`worlds' in our Everett theory. Moreover, the basic postulate of the neo-Everettian interpretation, that the state which can be decomposed is fully actualized, does not come from simply gazing at the mathematics, but from close argument that this is the best interpretation for quantum mechanics.

\subsubsection*{Error-correction}

The objection in Steane's fourth remark is from error-correction theory: the sensitivity to error in an $N$-qubit quantum computation is different from that of $2^N$ classical computations
running in parallel. Such a classical computer would be sensitive to errors of the order $1/2^N$, whereas from
error-correction theory we find the quantum computer to be sensitive only to O$(1/poly(N))$. Steane uses this to
question the idea that the $2^N$ calculations are actually taking place.

Again, this is explained naturally from neo-Everett. The difference with a classical parallel computer is that an error
process in a quantum computer (such as decoherence) will act on the whole state being processed. In other words, it
will act on all of the worlds identified within the state in exactly the same way. In classical parallel computing
errors can happen to individual computations (`worlds'), but because the worlds are not fully independent (see
above) in quantum computation, errors act globally. From this, we would \emph{expect} the sensitivity to error to
be O$(1/poly(N))$ -- it is not a surprise.

\subsubsection*{The Identification of Worlds}\label{www}


Remark 7, and the first part of section 3, deal with perceived problems of identifying worlds in the calculation in order to say that we have used `many worlds'. The point is first made that the different worlds are not fully independent --- evolution is unitary.
This is not, however, a problem for neo-Everett as the claim is only that the worlds are identifiable, insofar as we consider the
manipulation stage. That is all that is claimed, not that they are completely independent. The claim is that we can
\emph{identify} worlds within the state of the computation, rather than that the state is composed of many worlds. We will return to this point later.

Further, it is pointed out that the state of the system in computation is a single
entity, not a composition of many entities. Steane contends that it is a way of representing all the different states in a decomposition without `actually' calculating them or having them really exist. An analogy is drawn between
identifying worlds in the state and calling them real, and granting reality to the individual Fourier components of
a wave.

The idea that the state `represents' all the states of a decomposition without them being real is trying to show
that a many-worlds theory is not the only physical explanation of quantum computation, so is not relevant here.
The analogy with Fourier components is quite interesting, however, as it can be a classical example of what we are doing when we identify worlds in a state.

We note first of all that when we are talking about physical systems (rather than mathematical idealizations), `a
wave' is itself very much a structure defined `for all practical purposes'. It is the excitation of various parts of the medium (say water) in different ways at different times
--- yet we call it a single `thing' because it is relatively stable and acts independently and is explanatorily
useful (we can have a useful theory which talks about waves as single objects). We can mathematically analyze this
object in terms of its Fourier components. This is often useful mathematically; however in deciding whether or not
to grant them \emph{physical} reality, we must look at their physical usefulness. 

In some cases, \emph{pace} Steane, we do in fact wish to grant reality to the individual Fourier components, for exactly the same reasons that we wish to grant reality to individual worlds. One excellent example, given in \cite[p393]{toretti}, is the use of Fourier components in telecommunications:
\begin{quote} Telephone companies literally superpose the electromagnetic renderings of many simultaneous long-distance messages in a single wave train that is echoed by a satellite and then automatically analyzed at the destination exchange into its several components, each of which is transmitted over a separate private line. No doubt we may speak in this case of genuine spitting... the signal could also be split into other, meaningless components if the analysis were not guided by human interests and aims \end{quote}

Here we have re-enforced the point that the splitting of the world is not fundamental, but explanatory.



These points are very similar to an objection that is often raised to a picture of quantum computation in terms of many worlds \cite{jozsalecture}. Although it should be fairly evident by now that it does not impact on neo-Everett, it is worth dealing with in detail
to bring out one important aspect of the picture: the breakdown in some situations of the `worlds' concept. The objection
is that by the end of the computation we cannot tell in which of the computational worlds the computer
has been --- so why do we want to say that the `worlds' are individual and separate?

Even asking the question in this way does not make sense in neo-Everett. The state of the computation contains all the
different worlds --- this is the whole point, if it had only contained one then we would not have got speed up.
This is all, remember, within a single branch, or macrorealm: within the state of the quantum computer in our
single branch we can identify many computational worlds. As we have noted many times, the worlds are not wholly
independent, and cease even for practical purposes to be independent when they interfere.

This last point is the important one: \emph{at some points during the computation we can identify worlds
within the state, and at others we cannot}. That is, the `worlds' concept, as with all emergent concepts, breaks
down at some point\footnote{A good analogy with phonons is given in \cite{davidstr}: the concept of a phonon as
an entity is a good and useful one when they decay slowly on relevant time-scales, and once the decay becomes quick
the `phonon' concept begins to break down.}. This is neither a worry nor a problem, but simply a part of the fact
that we are dealing with structures that are only defined in certain practical situations. Consider a quantum Fourier transform-based algorithm, such as the Deutsch algorithm \cite{qft}. We can identify worlds from after the first Hadamard until after the manipulation stage, but in the
final Hadamard transformation the worlds concept breaks down as they interfere. All we can say is that before the
transformation we could identify $p$ worlds, and after we identified $q$ worlds. The worlds do not persist
throughout the entire computation. At various points we may describe the state of the computation in terms of many worlds, but there is nothing that requires that these worlds persist throughout the entire calculation -- just long enough for us to put them to use. This is, in fact, the same situation that we have even after decoherence has rendered worlds much more stable: it is still, in principle, possible to interfere worlds that have previously decohered. After that recoherence, there is no way of identifying `the' world that is `the' past history of an object. This is not just an experimental limitation: as it is an emergent concept this question simply makes no sense within the neo-Everettian notion of world.

\subsubsection*{Cluster-state computing}

We finish by looking at Steane's fifth remark. This concerns what is variously known as `measurement-based', `one-way', or `cluster-state' quantum computing \cite{dan}. The claim is that while a Fourier-transform based model of computing \emph{may} be amenable to a many-worlds description, this type of computing is not. In measurement-based quantum computing, a `cluster state' is prepared before the algorithm is chosen, containing all the entanglement that will be used during the computation. The algorithm itself is implemented by various stages of measurements on different parts of the cluster, usually with the measurements being dependent on the `feed-forward' of previous results. 

The Everettian description of this form of computing is in fact an area of active research, and points to what is probably the main outcome of the consideration of Everettian computing: quantum computing is not parallel classical computing\footnote{A good discussion of the problems inherent in a `parallel computing' description of quantum computing can be found in \cite[\S4.1]{chrisphilqi}.}. Computational worlds, if they can be described at all, are not fundamental to the ability of a quantum computer to out-perform its classical counterparts\footnote{The observant reader may note the divergence here from the view put forward in a previous manifestation of this paper (arXiv:quant-ph/0210204).}. While this goes against the usual informal view of many-worlds quantum computation, it is nevertheless an unavoidable conclusion of the neo-Everettian view. The computational ability of quantum computers in neo-Everett comes from elsewhere: the existence of the full quantum state, rather than one part of it. There seems, therefore, no \emph{prima facie} reason why the fully-quantum complete-state realism of neo-Everett will not be able to describe measurement-based computation usefully, as we do not need first to identify worlds in order to discuss the physical situation. We would, however, expect that the neo-Everettian picture will be quite different from a standard analysis of this type of computing, as such a picture would treat the measurements and feed-forward within the computation as entirely quantum mechanical. This would change our understanding of what is going on quite radically. Initial results support this conclusion, with the main mechanism for the computation being quantum information and entanglement transmitted through the supposedly `classical' channel of the measurement results. More work on the exact mechanisms, however, remains to be done.

\subsection*{`Full-state realism'?}

Mention of entanglement brings us to an interesting rapprochement between the neo-Everettian view and Steane's conclusions. Recent work on entanglement and computation \cite{mevent} demonstrates the fundamental physical role of entanglement in a computation considered from a neo-Everettian standpoint. Steane, in turn, concludes that entanglement rather than worlds leads to the quantum speed-up. Throughout his paper, Steane is absolutely right to reject a view of quantum computation that gives fundamental physical significance to computational worlds. This view is rejected in the neo-Everettian view as well. What we can see from the above discussion of Steane's paper is that this is a vitally important point. The interesting implications of the neo-Everettian view for computation do not depend so much on the idea that there are `many worlds', as on the notions of the existence of the complete state, no fundamental quantum/classical divide, and locality. 

This is in fact a wider issue: all too often when discussing the Everett interpretation, too much attention is focussed on the ``many worlds" part, frequently distracting from the real point that the worlds are not fundamental. One might almost be tempted to re-name the interpretation along the lines of ``full-state realism" to make this point: the identification of worlds within the state is often possible, and is useful when we wish to make contact between the quantum formalism and everyday experience (ie. to recover predictive power), but is not the basic physical picture given by the interpretation. The physical existence of the entire state in neo-Everett is what gives it its explanatory power, especially in quantum computation, where the `worlds' concept can break down\footnote{A further consequence of this is that, if this is so, then we may indeed wish to question our adoption of the language of many worlds in the case of coherent states, and follow the standard neo-Everettian line by keeping that to describe the situation post-decoherence.}. We need worlds for the explanatory power of the neo-Everettian interpretation as a whole (giving us the appearance of singular cats and boxes), but for computation it is the realism of the full state that gives us the physical picture.

There may, however, be one final twist for neo-Everettian quantum computing, arising from some very new work. We may be able to recover utility for a coherent worlds-based picture of computation in a framework where entanglement and superposition are the same behaviour manifesting between systems and worlds respectively \cite{vlatkoentsuper}. This work is, however, highly controversial, and the application to a many-worlds picture remains a conjecture -- albeit a tantalising one.

\section*{Summary and conclusions}

In this paper we have introduced the main ideas behind a neo-Everettian many-worlds interpretation of quantum mechanics, and its implications for quantum computation. We have discussed the general r\^{o}le of an interpretation in quantum theory, and seen how the neo-Everettian interpretation answers the questions raised. We have seen how the interpretation may be considered the best available interpretation, for its simplicity, explanatory power and its locality. We have also shown that the traditional problems raised against many-worlds theories do not impact upon it. 

In the field of quantum computation, we have seen how the main impact of the interpretation concerns locality, the lack of a sharp quantum/classical divide, and complete realism with respect to the state vector. This has implications both for how we describe a computation itself, and for the relationship between quantum and classical computations. Work on the neo-Everettian picture of computation is still ongoing, and it is likely that future developments will be able to clarify the r\^{o}le played by the concept of a `world' in coherent computing. A separate area of further research concerns new forms of computation. Discussion in the foundations of computation tends to concentrate on algorithms based on the quantum Fourier transform (QFT) as these were historically prior to other forms. The quantum walk form of computation \cite{viv} has not been considered in detail, yet has been an important form of computation for many years. It is foundationally interesting as, by contrast with QFT algorithms, there is no input during the computation: the algorithm is specified by the initial configuration. More recently, measurement-based quantum computing has thrown up some very interesting lines of research. This occupies a foundational area somewhere between QFT-based and quantum walk computing, as the initial state provides all the entanglement needed, but in order to run the algorithm there need to be a series of measurements on this state. We can see that this topic will continue to develop in interesting ways for some time to come.

\section*{Acknowledgements}

I gratefully acknowledge the hospitality and support of the Quinfo group at University College London during the writing of this paper, and I would also like to thank them for talk feedback which prompted me to revisit the discussion of locality. Thanks are due to the Quantum Information Group at the University of Leeds for their detailed response to a talk I gave, which gave rise to the pedagogy of this revised paper. I also acknowledge fruitful discussions on various elements of this paper, both in its present and past incarnations, especially with Vlatko Vedral, Harvey Brown, Hilary Cartaret, Viv Kendon, Chris Timpson, David Wallace, Andrew Steane, David Deutsch and Tom Boness. My thanks to all.

\newpage
\section*{Suggestions for further reading} 


There is a vast quantity of literature covering the topics touched on in this paper, and the following selection is neither complete nor comprehensive. The aim of this section is to suggest ways in which the interested reader can find out more about the various topics, and about the wider context in which this article can be placed. All the suggested articles in this section are available as preprints from either xxx.arxiv.org or philsci-archive.pitt.edu (and frequently both).

\subsection*{Interpreting quantum mechanics}

In this first section, I discuss two related areas within the philosophy of science: \emph{realism} and \emph{theory choice}. Realism is the idea that there is an external world about which we can learn and talk, and is a theory about what our scientific theories refer to. This contrasts with for example, \emph{empiricism}, which holds that, despite their external form, scientific theories are really just talking about the results of experiments. A good introduction to these ideas is the collection of articles\\

David Papineau. (ed.) \emph{The Philosophy of Science}, OUP 1996.\\

\noindent Also useful are the following:\\

James Ladyman. \emph{Understanding Philosophy of Science} Chs.5\&6, Routledge 2002.\\
\indent W. H. Newton-Smith. \emph{The Rationality of Science} Ch.II, Routledge 1981.

\noindent The debate about how we chose between theories has a long history to it. Chapters 1-4 of the above Ladyman book are good for this background. The solution presented here, that we chose the theory that is the best explanation of the observed phenomena, is known as \emph{inference to the best explanation} (IBE). This was presented in the classic text\\

Peter Lipton. \emph{Inference to the Best Explanation (Second Edition)}, Routledge 2004.\\

\noindent A helpful collection of texts for working out what exactly an explanation is, is\\

David-Hillel Ruben (ed.). \emph{Explanation}, Oxford Readings in Philosophy, OUP 1993.\\

\noindent Both realism and IBE are introduced informally by David Deutsch in his book\\

 David Deutsch. \emph{The Fabric of Reality} Chs.3\&7, Penguin Press 1997.\\

We have also, in this section, made use of the \emph{correspondence theory of truth}, sometimes known as \emph{Tarski truth} \cite{tarski}. This is the notion that a proposition X is true if and only if it is the case that X. This notion of truth is again part of the wider field of the philosophy of science, and not restricted to quantum theory. Tarski's formula was famously propounded in a science context by Popper:\\

Karl Popper. \emph{Conjectures and Refutations: The Growth of Scientific Knowledge}, Routledge 1963.


\subsection*{Many worlds}

Much has been written about the problem of measurement in quantum mechanics, and how the theory is to be interpreted (or not). A good introductory text is\\

David Albert. \emph{Quantum Mechanics and Experience}, Harvard University press 1994.\\

\noindent Another, despite its somewhat off-putting title, is\\

E. Squires. \emph{Conscious Mind in the Physical World}, Adam Hilger 1990.\\

\noindent The many-worlds and neo-Everettian interpretations are referenced fully in this paper, and the above books also contain introductions to general many-worlds type theories. Deutsch's \emph{Fabric of Reality} is again good for an informal introduction, although the reader should be aware that there are significant differences between his views and those of the neo-Everettian theory (for example, neo-Everett does not hold that quantum interference is physical scattering between systems in different worlds).


\subsection*{The incredulous stare}

The ideas of \emph{ontological simplicity} and \emph{structural simplicity} emerge frequently in discussions about theory choice. Ontological simplicity means that we are committed in a theory to less `stuff', and to fewer types of it. So, for example, a theory with normal things in it is ontologically simpler than a theory with normal things plus an aether. Structural simplicity means that the theory is less complex -- so, for example, the heliocentric model of planetary orbits is structurally simpler than the geocentric model which required epicycles etc. These two requirements will often be antagonistic. Ockham's Razor refers to ontological simplicity, and so is not the whole story. The online Stanford Encyclopedia has a good introductory article on the concepts of simplicity:\\

Alan Baker. http://plato.stanford.edu/entries/simplicity/ \\

\noindent Another good overview is\\

Elliott Sober. Simplicity, in W. H. Newton-Smith (ed.) \emph{A Companion to the Philosophy of Science} p433, Blackwell Publishing 2000.

\subsection*{Probability}

The problem of how we are to understand probability, in everyday life as well as in science, requires far more space to do it justice than was available here. In the discussion, I have concentrated mainly on they form of probability known as \emph{chance}, which is the type that is `out there' in the world, demonstrated by the decay of an atom or the fall of a coin. A full account of probability needs at the very least to consider the other types, which can generally be grouped as `credences' (what we believe to be the case, often explained in terms of how much we would bet on a given outcome) and `epistemic probabilities' (given all the evidence we have, what was the probability of a given event X). In this section I dealt briefly with the idea of probabilities being given meaning by ensembles of outcomes. This is known as the \emph{frequentist} theory of probability, and is one of several that have been offered to explain what we mean when we talk of probabilities.\\

\noindent A good and thorough text for understanding the arguments around probability is\\

D. H. Mellor. \emph{Probability -- A Philosophical Introduction}, Routledge 2005.\\

\noindent For the Deutsch-Wallace decision theory programme to use the Everett interpretation to give meaning to probabilities, see\\

David Deutsch. Quantum theory of probability and decisions, \emph{Proceedings of the Royal Society of London} A455, 3129-37, 1999.\\

David Wallace. Everettian rationality: defending Deutsch's approach to probability in the Everett interpretation, \emph{Studies in History and Philosophy of Modern Physics} 34, 415-439, 2003.\\

Hilary Greaves. Probability in the Everett interpretation, \emph{Philosophy Compass} 2(1), 109–128, 2007.\\

\noindent For some of the objections to the programme, see

Huw Price. Decisions, decisions, decisions: can Savage salvage Everettian probability? \emph{Pittsburg e-print} 00003886, 2008.\\


\subsection*{Locality}

Questions of locality and nonlocality in quantum theory have given physicists and philosophers of physics many headaches over the years. It is important to realize precisely what is and is not shown by the EPR and Bell papers, and the Aspect experiments. A very good and thorough introduction to the area is given by\\

Tim Maudlin. \emph{Quantum Non-Locality and Relativity (Second Edition)}, Blackwell Publishing 2002.\\

\noindent Another good introduction is\\

Marc Lange. \emph{An Introduction to the Philosophy of Physics} Ch.9, Blackwell Publishing 2002.\\

\noindent For a detailed account of the differing types of locality used in the Bell theorems, and the possible connections between quantum mechanical (non)locality and relativity, see\\

Jeremy Butterfield. Stochastic Einstein locality revisited, \emph{Brit J Philos Sci} 2007; 58: 805 - 867.


\subsection*{Quantum computing}


The implications of quantum computing and information theory for the foundations of physics (and \emph{vice versa}) is a small but active topic of ongoing research. A very good article on this area is\\

Chris Timpson. Philosophical aspects of quantum information theory in D. Rickles (ed.) \emph{The Ashgate Companion to the New Philosophy of Physics} Ashgate 2008 (arXiv:quant-ph/0611187).\\

\noindent A useful resource for articles on this topic is the special `Quantum Information' journal edition\\

Volume 34, number 3 of \emph{Studies in History and Philosophy of Modern Physics}, 2003.

\newpage
\bibliographystyle{nature}
\bibliography{biblogMAIN}

\end{document}